\documentclass[%
12pt,
preprint,
superscriptaddress,
preprintnumbers,
amsmath,amssymb,
aps,
prb,
floatfix,
]{revtex4-2}

\usepackage{graphicx}% Include figure files
\usepackage{rotating}
\usepackage{dcolumn}% Align table columns on the decimal point
\usepackage{bm}% bold math
\usepackage[colorlinks, allcolors=blue]{hyperref}% add hypertext capabilities
\usepackage[mathlines]{lineno}% Enable numbering of text and display math
\usepackage{orcidlink}
\usepackage{sidecap}
\usepackage{dcolumn}
\usepackage[utf8]{inputenc}

\usepackage{xcolor}
\usepackage{microtype}
\usepackage{graphicx}
\usepackage[multidot]{grffile}
\graphicspath{ {figs/} } 

\usepackage[subpreambles=true]{standalone}

\newcommand{\pbt}{P\bar{3}}
\newcommand{\pbtm}{P\bar{3}m1}
\newcommand{\rscan}{$r^2$SCAN } %one blank is vital

\begin{document}

\title{\textit{Ab initio} study on the stability and elasticity of brucite}
\author{Hongjin Wang\,\orcidlink{0009-0008-9766-1177}}
\affiliation{Department of Applied Physics and Applied Mathematics, Columbia University, New York, New York 10027, USA}
\author{Chenxing Luo\,\orcidlink{0000-0003-4116-6851}}
\affiliation{Department of Applied Physics and Applied Mathematics, Columbia University, New York, New York 10027, USA}
\author{Renata M. Wentzcovitch\,\orcidlink{0000-0001-5663-9426}}
\email[]{rmw2150@columbia.edu}
\affiliation{Department of Applied Physics and Applied Mathematics, Columbia University, New York, New York 10027, USA}
\affiliation{Department of Earth and Environmental Sciences, Columbia University, New York, New York 10027, USA}
\affiliation{Lamont--Doherty Earth Observatory, Columbia University, Palisades, New York 10964, USA}
\affiliation{Data Science Institute, Columbia University, New York, NY 10027, USA}

\date{\today}

\begin{abstract}
Brucite (Mg(OH)$_2$) is a mineral of great interest owing to its various applications and roles in geological processes.
Its structure, behavior under different conditions, and unique properties have been the subject of numerous studies and persistent debate.
As a stable hydrous phase in subduction zones, its elastic anisotropy can significantly contribute to the seismological properties of these regions.
We performed \textit{ab initio} calculations to investigate brucite's stability, elasticity, and acoustic velocities.
We tested several exchange-correlation functionals and managed to obtain stable phonons for the $\pbt$ phase with \rscan for the first time at all relevant pressures up to the mantle transition zone.
We show that \rscan performs very well in brucite, reproducing the experimental equation of state and several key structure parameters related to hydrogen positions.
The room temperature elasticity results in $\pbt$ reproduces the experimental results at ambient pressure.
These results, together with the stable phonon dispersion of $\pbt$ at all relevant pressures, indicate $\pbt$ is the stable candidate phase not only at elevated pressures but also at ambient conditions.
The success of \rscan in brucite, suggests this functional should be suitable for other challenging layer-structured minerals, e.g., serpentines, of great geophysical significance.
\end{abstract}

\maketitle

\section{Introduction}

Water, incorporated stoichiometrically in hydrous minerals, as hydrous defects, or in melts, plays a pivotal role in Earth's interior dynamics.
It lowers the sub-solidus viscosity of rocks and internal frictions facilitating earthquakes \cite{ohtani_role_2020,faccenda_water_2014}.
Each year, significant amounts of water ($\sim$$10^{11}$~kg) enter Earth's mantle through the subduction process \cite{vankekenSubductionFactoryDepthdependent2011}.
In the upper mantle, water reacts with rocks beneath the crust to form hydrous minerals and enters the subduction process.
Carried to greater depths, increased pressure and temperature produce dehydration, resulting in slab embrittlement.
Formed at the early stages of the subduction process, sheet-like hydrous minerals, e.g., brucite and serpentine phases, are the major water carriers at upper mantle depths.
These minerals consist of Mg-Si-O-containing layers bounded only by hydrogen bonds (H-bonds) and are weak in shear strengths, facilitating the slab subduction process \cite{evans_serpentinite_2004,evans_serpentinite_2013}.
Given the intricate details of their formation and participation in geotectonic processes, the understanding of serpentine phases deepens our grasp of geological processes in the upper mantle and the importance of water in shaping them.

To address phase relations in these hydrous phases, it is necessary to address brucite (Mg(OH)$_2$) first.
As one of the simplest and most water-abundant minerals within the ternary MgO-SiO$_2$-H$_2$O (MSH) system \cite{hermann_high-pressure_2016},
brucite is central in the many geochemical processes of serpentine formation and transformation that occur in subducted slabs.
For example, brucite participates in the retrograde formation of antigorite from olivine (Mg$_{2}$SiO$_4$) or dehydration of antigorite (Mg$_{3m-3}$Si$_{2m}$O$_{5m}$(OH)$_{4m-6}$) \cite{evans_serpentinite_2004} and in the lizardite (Mg$_3$Si$_2$O$_5$(OH)$_4$) to antigorite
transition \cite{evans_serpentinite_2004, Ghaderibrucite, dengElasticAnisotropyLizardite2022}, both being serpentine-type phases.
So, we need to address brucite also to understand phase relations between these complex serpentine phases.

Ever since the structure of brucite was first refined in 1967 with the space group $\pbtm$ \cite{Zigan1967NeutronenbeugungsmessungenAB}, such assignment has been a subject of continuous debate. 
The original work did not account for the ``riding" motion of hydrogen and got an inaccurate O---H bond length \cite{partinCrystalStructureProfile1994}. 
Neutron diffraction \cite{catti1995static,parise1994pressure} and \textit{ab initio} molecular dynamics simulations \cite{raugei1999prl} suggest protons are displaced from the threefold axis at elevated pressures.
To accommodate this disordered displacement of protons from high-symmetry positions, an enlarged $\sqrt{3}\times\sqrt{3}\times1$ supercell structure with lower, $\pbt$ symmetry, was proposed in 2006 \cite{mookherjee2006high}.
Still, no consensus has been reached on whether $\pbtm$, \cite{pillai2021brucite,pishtshevMaterialsPropertiesMagnesium2014} or $\pbt$ \cite{jochym2010brucite}, or a structure with even lower symmetry \cite{trevino2018anharmonic} describes brucite's stable phase at ambient conditions.

Resolving the structure of brucite and being able to describe its vibrational properties is the first step to computing its thermodynamic and thermoelastic properties at high pressures and high temperatures (high-$PT$).
The Wu-Wentzcovitch semi-analytical method (SAM-Cij) \cite{wuQuasiharmonicThermalElasticity2011, luoCijPythonCode2021} is a concise and predictive formalism to calculate the thermoelastic tensor (Cij) of crystalline solids that depends on the existence of stable phonons \cite{sunDynamicStabilizationCubic2014b,WANGpgm,ZhuangFeprb}, still debatable for brucite minerals \cite{trevino2018anharmonic,pillai2021brucite}. This method has been successfully applied to materials across all crystal systems in conjunction with \textit{ab initio} calculations of static elastic coefficients and phonon frequencies. 

This work reevaluates brucite's phonon stability issue using density-functional theory (DFT) with \rscan\cite{r2scancitation} description of the exchange-correlation energy.
Compared to PBE \cite{pbecitation} and LDA \cite{LDAcitation}, the SCAN meta-GGA functionals \cite{sunStronglyConstrainedAppropriately2015} describes better the H-bond compressive behavior\cite{luoHighThroughputSampling2023, wanThermoelasticPropertiesBridgmanite2023, sunAccurateFirstprinciplesStructures2016}.
The \rscan functional has the same accuracy as the SCAN functional but performs better numerically. 
It also slightly improves dynamic stability \cite{ning_reliable_2022}.
This method stabilizes the $P\bar{3}$ structure's phonon dispersion in a wide pressure range of geophysical significance.
Here we report the high-temperature elastic properties and velocities of brucite, for the first time, using the SAM-Cij method.

%Some introduction about Brucite. The debate on which phase is more stable. $P\bar{3}m1$ or $P\bar{3}$. Also a lot of debate on which phase has stable phonons.

This paper is organized as follows.
Section~\ref{sec:method} introduces the method and DFT calculation parameters.
We show results and compare them with measurements and previous calculations in Section~\ref{sec:result}.
Summary and conclusions are presented in Section~\ref{sec:conclusion}.

\section{Method}
\label{sec:method}

All density-functional theory calculations were conducted using the projector augmented wave (PAW) method \cite{PAWcitation} implemented in the VASP code \cite{vaspcitation}. %on $\sqrt{3}$ × $\sqrt{3}$ × 1 supercells for $P\bar{3}$ phase brucite. 
We employed the \rscan \cite{r2scancitation} exchange-correlation functional.
Local density approximations (LDA) \cite{LDAcitation} and generalized gradient approximations (GGA) parameterized by the Perdew-Burke-Ernzerhof formula (PBE) \cite{pbecitation} were also used to compare with the \rscan results.
A plane-wave basis was used with a kinetic energy cutoff of 800~eV.
We chose a $k$-point of 4 × 4 × 4, which ensures the calculations are converged to $10^{-8}$~eV.
The full Brillouin zone phonon spectra were calculated on the 2 × 2 × 2 supercells using the finite displacement method implemented in \textsc{Phonopy} \cite{phonopy-phono3py-JPCM}.
The structures at elevated pressures are calculated by uniformly scaling the structure at ambient pressure and then relaxing at constant volumes.
The static elastic coefficient tensor was obtained using the stress vs.\ strain relations resulting from $\pm$1\% infinitesimal strain.

The high-temperature elastic tensor is calculated using the Python package \texttt{cij} \cite{luoCijPythonCode2021}.
Its components are written as a derivative of the Helmholtz free energy under isothermal conditions \cite{barron1965second, wuQuasiharmonicThermalElasticity2011, luoCijPythonCode2021}:
\begin{equation}
c^T_{ijkl}=\frac{1}{V}\left(\frac{\partial^2 F}{\partial e_{ij} \partial e_{kl}}\right)+\frac{1}{2}P(2\delta_{ij}\delta_{kl}-\delta_{il}\delta_{kj}-\delta_{ik}\delta_{jl})
\end{equation}
Here, $e_{ij},e_{kl}$~($i,j,k,l=1,2,3$) represents the infinitesimal strains, $P$ represents pressure, $\delta_{ij}$ represents the Kronecker delta symbol, and $F$ represents the Helmholtz free energy, which can be calculated using the quasiharmonic approximation (QHA) \cite{qinQhaPythonPackage2019}:
\begin{equation}
\begin{split}
    F(e_{ij}, V, T)=U(e_{ij}, V) + \frac{1}{2}\sum_{q,m}\hbar \omega_{q,m}(e_{ij},V)\\
    +\,k_B T\sum_{q,m}\ln\left\{1-\exp \left[- \frac{\hbar \omega_{q,m}(e_{ij},V)}{k_B T}\right]\right\}.
\end{split}
\end{equation}
Here, $\omega_{q,m}$ represents the vibrational frequency of the $m$-th normal mode with the phonon wave vector $q$. $V$, $T$ represent the equilibrium volume and temperature. 
$\hbar$ and $k_B$ are the reduced Planck and Boltzmann constant, respectively.

Crystal structure images are created using VESTA \cite{vestacitation}.

\section{Results and Discussion}
\label{sec:result}

\subsection{Structure optimization}
\begin{figure}[htbp]
\centering
\includegraphics[width=0.5\textwidth]{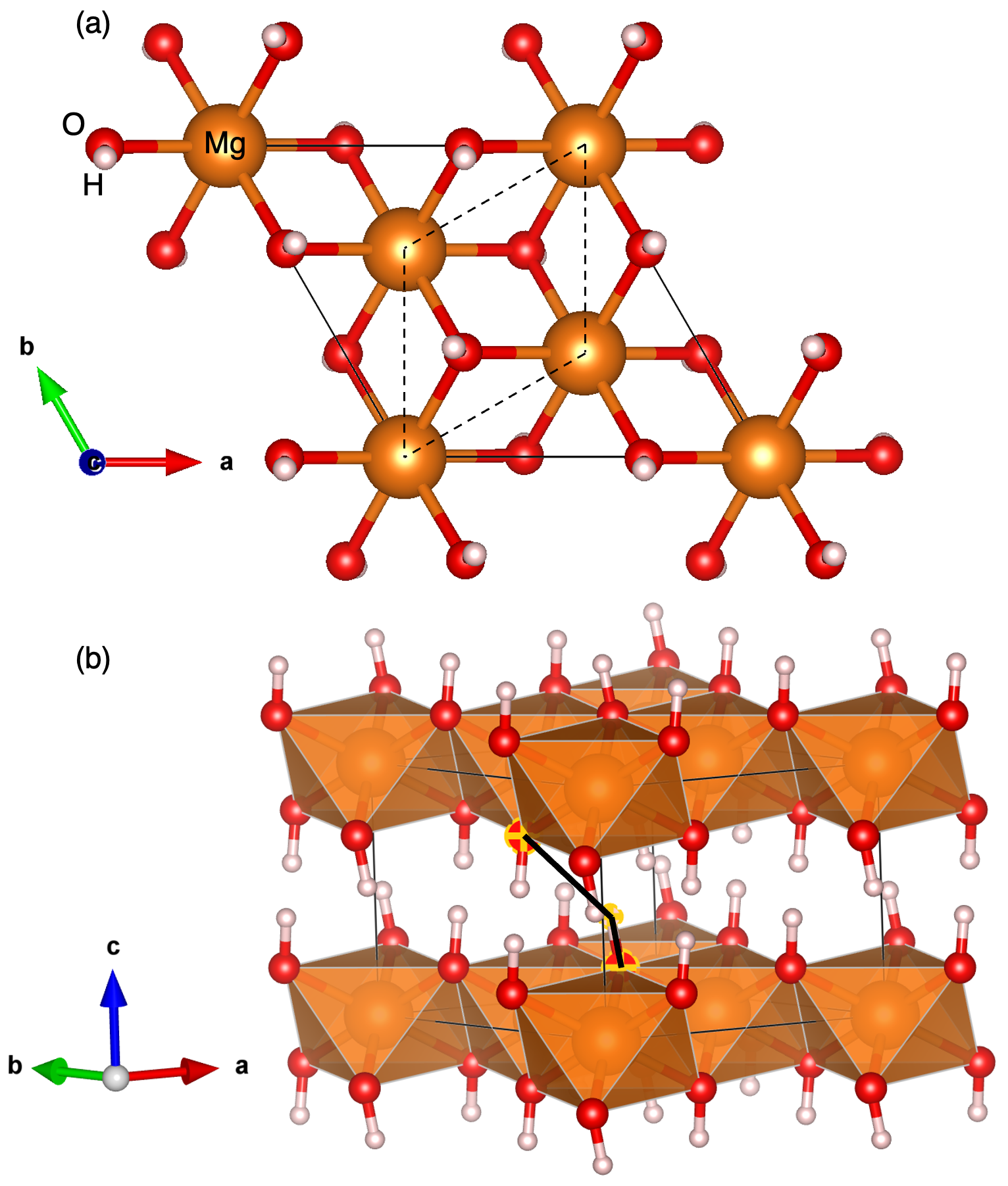}
\caption{Crystal structure of brucite: (a)~Top view of the unit cell of a $\pbt$ brucite with lattice vectors shown as solid black lines. The lattice vectors of the smaller and more symmetric $\pbtm$ structure are shown as dashed lines. (b)~Side view of a $\pbt$ brucite. Highlighted atoms show a typical interlayer H-bond (bold solid lines). Yellow: Mg; red: O; white: H}
\label{fig:structure}
\end{figure}
The crystal structure of brucite was first refined by Zigan and Rothbauer \cite{Zigan1967NeutronenbeugungsmessungenAB} in the space group $P\bar{3}m1$ with protons on the threefold axis corresponding to the $2d$ Wyckoff position $(1/3, 2/3, z)$.
Each hydroxyl is surrounded by three oppositely oriented hydroxyl groups of the overlying layer.
Subsequent theoretical \cite{raugei1999prl} and experimental studies \cite{desgranges1996interlayer} found the protons tend to displace from the $2d$ site to the $6i$ Wyckoff site at $(x, 2x, z)$ with an occupation of 1/3 at elevated pressure.
The value of $x$ could be either $> 1/3$ (so-called the $x$ greater than third, or XGT, arrangement) or $< 1/3$ (the $x$ less than third, or XLT, arrangement).
In \textit{ab initio} calculations, the 1/3 occupation can be simulated by supercells.
Mookerjee and Stixrude \cite{mookherjee2006high} modeled the transition of hydrogen to XGT position using an enlarged $\sqrt{3}\times\sqrt{3}\times1$ supercell. This model minimizes the interlayer H$\cdots$O distance, resulting in a $P\bar{3}$ symmetry (a maximum subgroup of $P\bar{3}m1$) for brucite.
The XLT arrangement conserves the $\pbtm$ symmetry but is energetically unfavorable.
It is important to note that the $\pbt$ model is equivalent to $\pbtm$ with displaced hydrogen at the $6i$ position.
The relation of $\pbt$ and $\pbtm$ brucite is shown in Figure \ref{fig:structure}.

Table~\ref{structure_table} shows the lattice parameters and volumes at ambient conditions compared to the previous calculations and measurements.
In $\pbt$, $a$ is divided by $\sqrt{3}$ for comparison with $\pbtm$.
All experiments in Table~\ref{structure_table} are at room temperature and ambient pressure, and all calculations are static 0~GPa unless otherwise stated.
It is also important to notice that even though all experiments assume the space group of $\pbtm$, some did obtain the results where hydrogen atoms are favored at $6i$ positions \cite{parise1994pressure,desgranges1996interlayer,catti1995static}.
The most important factor for all calculations is the choice of exchange-correlation functionals. 
Our results are similar to previous results with the same exchange-correlation functionals.
The $\pbt$ brucite is slightly smaller than the $\pbtm$ counterpart.
The difference between the two phases also depends heavily on the choice of exchange-correlation functionals.
For PBE or PBEsol functionals, the difference is minor, at 0.001~{\AA} for $a$ and 0.01~{\AA} for $c$.
Our LDA results show a difference of 0.02~{\AA} in $a$ and 0.1~{\AA} in $c$. 
Also, since all measurements are performed at room temperature, an ideal description of static volume needs to be smaller than the measurements (volume $< 41$~{\AA}$^3$).
In Table~\ref{structure_table}, only LDA, B3YLP-D, and $r^2$SCAN satisfy this condition.
In addition, Ref.~\cite{Ghaderibrucite} showed the LDA results with 300~K thermal correction, and the results still differ considerably from the experimental results.
From the static results, $r^2$SCAN and B3YLP are better choices for the brucite system than LDA and PBE, but LDA and PBE provide good lower and upper bounds for the key structure parameters.

\begin{table*}
\caption{Key structure parameters data compared to previous calculations and measurements. All experiments are at room temperature and ambient pressure, and all calculations are static at 0~GPa unless otherwise stated.}
\label{structure_table}
\begin{ruledtabular}
\begin{tabular}{ccccccc}
%\toprule
Space group & $a$ (\AA) & $c$ (\AA) & $c/a$ & Volume (\AA$^3$) & Method & Reference\\
\hline
$\pbtm$ & 3.146 & 4.768 & 1.515 & 40.868 & X-ray Diffraction & Fei and Mao (1993) \cite{feimaobrucite} \\
$\pbtm$ & 3.138 & 4.713 & 1.502 & 40.200 & Neutron Diffraction & Parise (1994) \cite{parise1994pressure} \\
$\pbtm$ & 3.150 & 4.720 & 1.498 & 40.560 & Neutron Diffraction & Catti (1995) \cite{catti1995static} \\
$\pbtm$ & 3.145 & 4.769 & 1.516 & 40.851 & X-ray Diffraction & Duffy (1995) \cite{duffy1995single} \\
$\pbtm$ & 3.148 & 4.779 & 1.518 & 41.015 & Neutron Diffraction & Desgranges (1996) \cite{desgranges1996interlayer} \\
$\pbtm$ & 3.148 & 4.771 & 1.516 & 40.930 & X-ray Diffraction & Fukui (2003) \cite{fukui2003brucite} \\
$\pbtm$ & 3.150 & 4.783 & 1.518 & 41.101 & Brillouin Scattering & Jiang (2006) \cite{jiang2006brucite} \\
$\pbtm$ & 3.155 & 4.772 & 1.512 & 41.147 & Neutron Diffraction & Xu (2013) \cite{xu2013high} \\
$\pbtm$ & 3.147 & 4.757 & 1.512 & 40.793 & X-ray Diffraction & Ma (2013) \cite{ma2013brucite} \\
$\pbtm$ & 3.146 & 4.770 & 1.516 & 40.890 & X-ray Diffraction & Pilai (2021) \cite{pillai2021brucite} \\
\hline
$\pbt$ & 3.199 & 4.844 & 1.514 & 41.780 & PBE & Moohkerjee and Stixrude (2006) \cite{mookherjee2006high} \\
$\pbt$ & 3.070 & 4.400 & 1.433 & 36.000 & LDA & Ghaderi (2015) \cite{Ghaderibrucite} \\
$\pbt$ & 3.080 & 4.440 & 1.442 & 36.660 & LDA+MGD~300~K & Ghaderi (2015) \cite{Ghaderibrucite} \\
$\pbt$ & 3.200 & 4.840 & 1.513 & 41.700 & PBE & Ghaderi (2015) \cite{Ghaderibrucite} \\
$\pbtm$ & 3.188 & 4.786 & 1.501 & 42.120 & PBEsol & Treviño (2018) \cite{trevino2018anharmonic} \\
$\pbt$ & 3.186 & 4.777 & 1.499 & 42.000 & PBEsol & Treviño (2018) \cite{trevino2018anharmonic} \\
$\pbtm$ & 3.141 & 4.666 & 1.486 & 39.900 & B3YLP-D & Ulian (2019) \cite{ulian2019brucite} \\
$\pbt$ & 3.129 & 4.657 & 1.488 & 39.500 & B3YLP-D & Ulian (2019) \cite{ulian2019brucite} \\
$\pbtm$ & 3.186 & 4.848 & 1.522 & 42.611 & vdW-DF2 & Pilai (2021) \cite{pillai2021brucite} \\
\hline
$\pbtm$ & 3.092 & 4.510 & 1.458 & 37.350 & LDA & This work \\
$\pbt$ & 3.074 & 4.418 & 1.437 & 36.164 & LDA & This work \\
$\pbtm$ & 3.183 & 4.900 & 1.539 & 42.997 & PBE & This work \\
$\pbt$ & 3.182 & 4.894 & 1.538 & 42.927 & PBE & This work \\
$\pbtm$ & 3.139 & 4.730 & 1.507 & 40.374 & $r^2$SCAN & This work \\
$\pbt$ & 3.129 & 4.662 & 1.490 & 39.532 & $r^2$SCAN & This work \\
%\bottomrule
\end{tabular}
\end{ruledtabular}
\end{table*}

\subsection{Phonon stability}
Phonon stability in brucite has been a debated topic in recent years.
Treviño et al.~\cite{trevino2018anharmonic} suggested that brucite needs to transform into a structure with an even lower symmetry $C2/m$ than $P\bar{3}$ or $P\bar{3}m1$.
They ruled out $P\bar{3}$ and $P\bar{3}m1$ phases as they display unstable temperature-dependent phonons at 300~K.
According to their 300~K calculations using PBEsol, only $C2/m$ has stable phonons; at higher temperatures, the other phases would also display stable phonons.
However, in their comparison, the $q$-path for the $C2/m$ phase dispersion is non-equivalent from those in $\pbtm$ and $\pbt$, which makes the comparison ambiguous.
Also, their choice of PBEsol overestimates the volume of both $\pbtm$ and $\pbt$ phases (see Table \ref{structure_table}).

\textit{Ab initio} phonon calculations performed by Pillai et al.~\cite{pillai2021brucite} suggested the $P\bar{3}m1$ symmetry would be the stable structure at ambient conditions using a non-empirical van der Waals functional (vdW-DF2) method.
The discrepancy in their results suggests the importance of suitable exchange-correlation functionals for brucite.
However, as shown in Table \ref{structure_table}, the vdW-DF2 results also overestimate the volume.

\begin{figure}[htbp]
\centering
\includegraphics[width=0.5\textwidth]{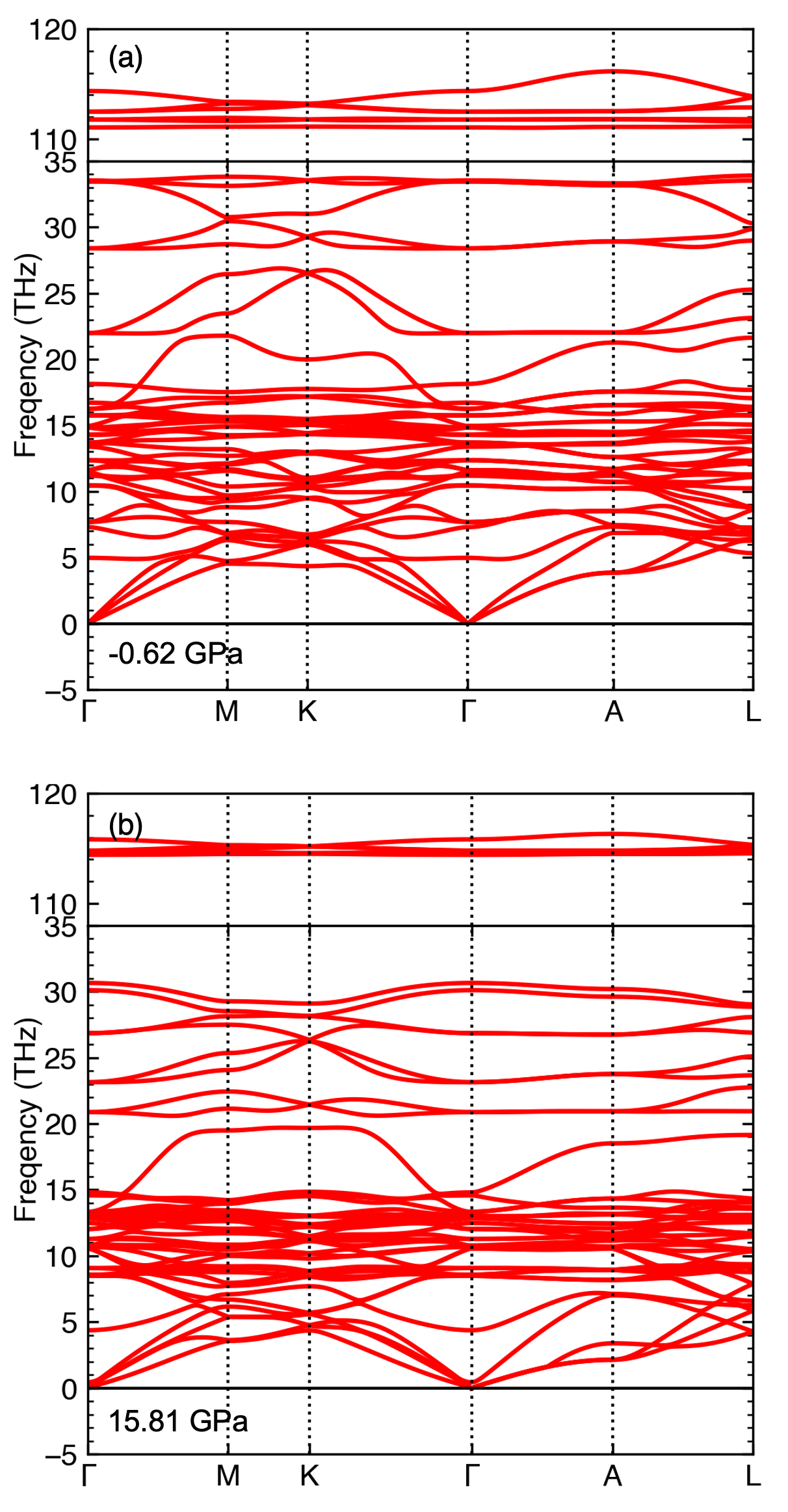}
\caption{Phonon spectra of $\pbt$ brucite at (a) ambient and (b) elevated pressure. In both cases, the phonon remains stable.}
\label{sample_phonon}
\end{figure}

The success of \rscan in describing the brucite structures suggests this functional should also be successful for phonon calculations.
We calculate phonon dispersions for $P\bar{3}m1$ and $P\bar{3}$ phases at a series of pressures.
Figure \ref{sample_phonon} shows the phonon dispersions of $\pbt$ brucite at ambient and elevated pressures.
Detailed phonon dispersions for $\pbtm$ and $\pbt$ brucite at several pressures can be found in the supplemental Figs. S1 and S2.
In contrast to previous results using other functionals \cite{trevino2018anharmonic,pillai2021brucite}, we find stable phonons for $P\bar{3}$ at all investigated pressures and unstable modes for $P\bar{3}m1$ (Figs. S1 and S2).
This indicates that \rscan, a functional that successfully describes H-bonds, can also stabilize $\pbt$ brucite at low and elevated pressures.
% {\color{red}, which indicates that ... is stable.}
Although the consensus seems to be that $\pbt$ is the stable brucite phase at higher pressure, this is the first time that stable phonons of $\pbt$ brucite are shown at elevated pressures.

\subsection{Equation of state (EoS)}
% EoS figure subject to change

\begin{figure}[htbp]
\centering
\includegraphics[scale=0.8]{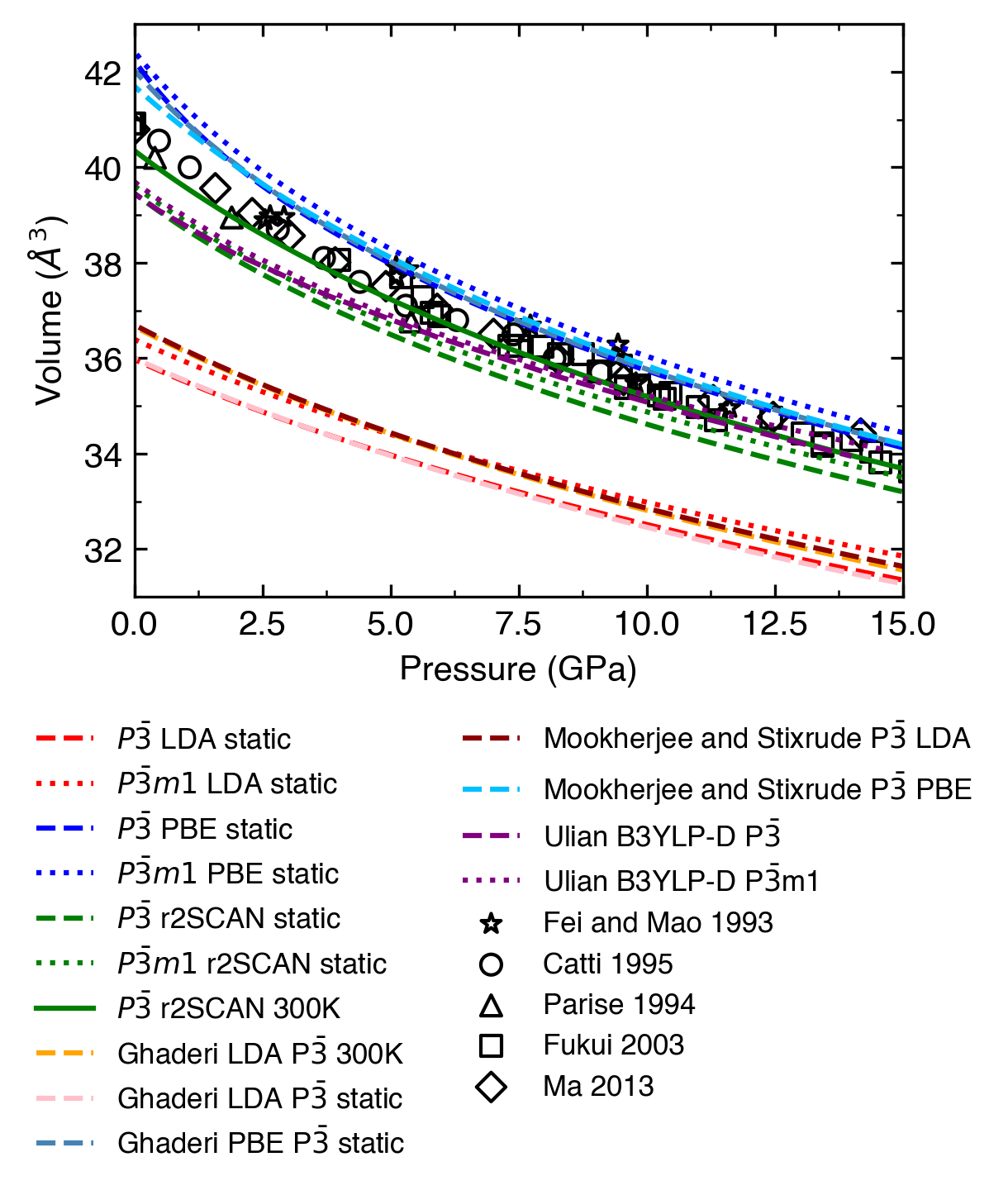}
\caption{Equation of state of $\pbt$ and $\pbtm$ brucite phases calculated with different exchange-correlation functionals. 
Dashed lines: $\pbt$; dotted lines: $\pbtm$. 
Different colors represent different exchange-correlation functionals (red-like: LDA; blue-like: PBE; green: \rscan; purple: B3YLP).}
\label{eos_fig}
\end{figure}

Fig.~\ref{eos_fig} shows our calculated equation of states using different exchange-correlation functionals and comparisons with measurements \cite{feimaobrucite,catti1995static,parise1994pressure,fukui2003brucite,ma2013brucite} and previous \textit{ab initio} results \cite{Ghaderibrucite,mookherjee2006high,ulian2019brucite}.
The static compression curve is obtained by fitting the free energy vs.\ volume with a third-order finite strain equation of state and then computing pressure as the free energy derivative w.r.t volume.
Our benchmark static LDA and PBE results agree well with previous calculations \cite{Ghaderibrucite,mookherjee2006high}, showing $\sim$10\% underestimation of volume by LDA and $\sim$5\% overestimation with PBE compared to the measurements.
\rscan and B3YLP-D \cite{ulian2019brucite} results lie between LDA and PBE bounds and, at high pressures, lean close to PBE results.
However, the semi-empirical B3YLP-D \cite{ulian2019brucite} functional does not seem to account for the temperature effect as the others, i.e., increase the volume uniformly from the static results, either using QHA or the Mie-Debye-Gr\"uneisen model \cite{mookherjee2006high,Ghaderibrucite}.
The B3YLP-D compression curve shape differs from the measurements also.
Accounting for thermal effects, the B3YLP-D functional should overestimate the volume at higher the high pressures of geophysical significance.
300~K QHA \rscan results are quite accurate between 0--5~GPa, underscoring the necessity of including the finite-temperature effects in all calculations.
Vibrational effects correct the volume by $\sim$5\% at all pressures.
The overall difference in volumes between 300~K QHA and measurements is less than 0.5~{\AA}$^3$, less than the difference between measurements.

The stable \rscan dispersions of $\pbt$ brucite can be used to calculate the finite-temperature equation of state using the QHA method.
Fitting $P$-$V$ data to third order Birch–Murnaghan EoS gives $V_0=40.35$~\AA$^3$, $B_0=49.35$~GPa, and $B^\prime=5.82$.
Table~\ref{tab:eos_table} compares our EoS fitting parameters with previous measurements \cite{feimaobrucite,parise1994pressure,catti1995static,duffy1995high,jiang2006brucite,ma2013brucite} and calculations \cite{mookherjee2006high,Ghaderibrucite,ulian2019brucite}.
Our $V_0$ is within the range of all measurements and agrees better with them than all previous calculations.
Similarly, our $B_0$ and $B^\prime$ are also within the range of measurements. However, different measurements vary considerably, which might stem from the different pressure ranges these measurements used.
Among all calculations, the fitting parameters from our \rscan 300~K results agree best with measurements.

\begin{table*}[h]
\centering
\caption{Equation of state parameters compared to previous calculations and measurements. All experiments are at room temperature and ambient pressure, and all calculations are static unless otherwise stated.}
\label{tab:eos_table}
\begin{ruledtabular}
\begin{tabular}{cccccc}
Space Group &  $V_0$  ({\AA}$^3$)   &  $B_0$ (GPa)     &  $B^\prime$    & Method         & Reference                \\ \hline
$\pbtm$       & 40.87               & 54.30            & 4.70     & X-ray Diffraction    & Fei and Mao (1993) \cite{feimaobrucite}        \\
$\pbtm$       & 40.20               & 47.00            & 4.70     & Neutron Diffraction  & Parise (1994) \cite{parise1994pressure}\\
$\pbtm$       & 40.99               & 39.00            & 7.60     & Neutron Diffraction  & Catti (1995) \cite{catti1995static}              \\
$\pbtm$       & 40.85               & 42.00            & 5.70     & X-ray Diffraction    & Duffy (1995) \cite{duffy1995single}              \\
$\pbtm$       & 40.93               & 41.80            & 6.60     & X-ray Diffraction    & Fukui (2003) \cite{fukui2003brucite}              \\
$\pbtm$       & 41.10               & 35.80            & 8.90     & Brillouin scattering & Jiang (2006) \cite{jiang2006brucite}              \\
$\pbtm$       & 40.79               & 37.00            & 10.60    & X-ray Diffraction    & Ma (2013) \cite{ma2013brucite}                 \\ \hline
$\pbt$         & 41.70               & 43.00            & 5.70     & PBE                  & Moohkerjee and Stixrude (2006) \cite{mookherjee2006high} \\
$\pbt$         & 36.70               & 65.00            & 6.05     & LDA                  & Moohkerjee and Stixrude (2006) \cite{mookherjee2006high} \\
$\pbt$         & 43.80               & 34.00            & 5.80     & PBE 300~K             & Moohkerjee and Stixrude (2006) \cite{mookherjee2006high} \\
$\pbt$         & 38.20               & 53.00            & 6.20     & LDA 300~K             & Moohkerjee and Stixrude (2006) \cite{mookherjee2006high} \\
$\pbt$         & 36.00               & 73.40            & 5.30     & LDA                  & Ghaderi (2015) \cite{Ghaderibrucite}             \\
$\pbt$         & 36.66               & 66.90            & 5.40     & LDA 300~K             & Ghaderi (2015) \cite{Ghaderibrucite}             \\
$\pbt$         & 42.03               & 34.60            & 7.50     & PBE                  & Ghaderi (2015) \cite{Ghaderibrucite}             \\
$\pbtm$       & 39.70               & 47.50            & 10.10    & B3YLP-D              & Ulian (2019) \cite{ulian2019brucite}               \\
$\pbt$         & 39.59               & 48.00            & 9.10     & B3YLP-D              & Ulian (2019) \cite{ulian2019brucite}               \\
$\pbt$         & 40.35               & 49.35            & 5.82     & \rscan 300~K          & This work                \\ 
\end{tabular}
\end{ruledtabular}
\end{table*}

\subsection{Compression behavior}
\begin{figure}[htbp]
\centering
\includegraphics[width=0.42\textwidth]{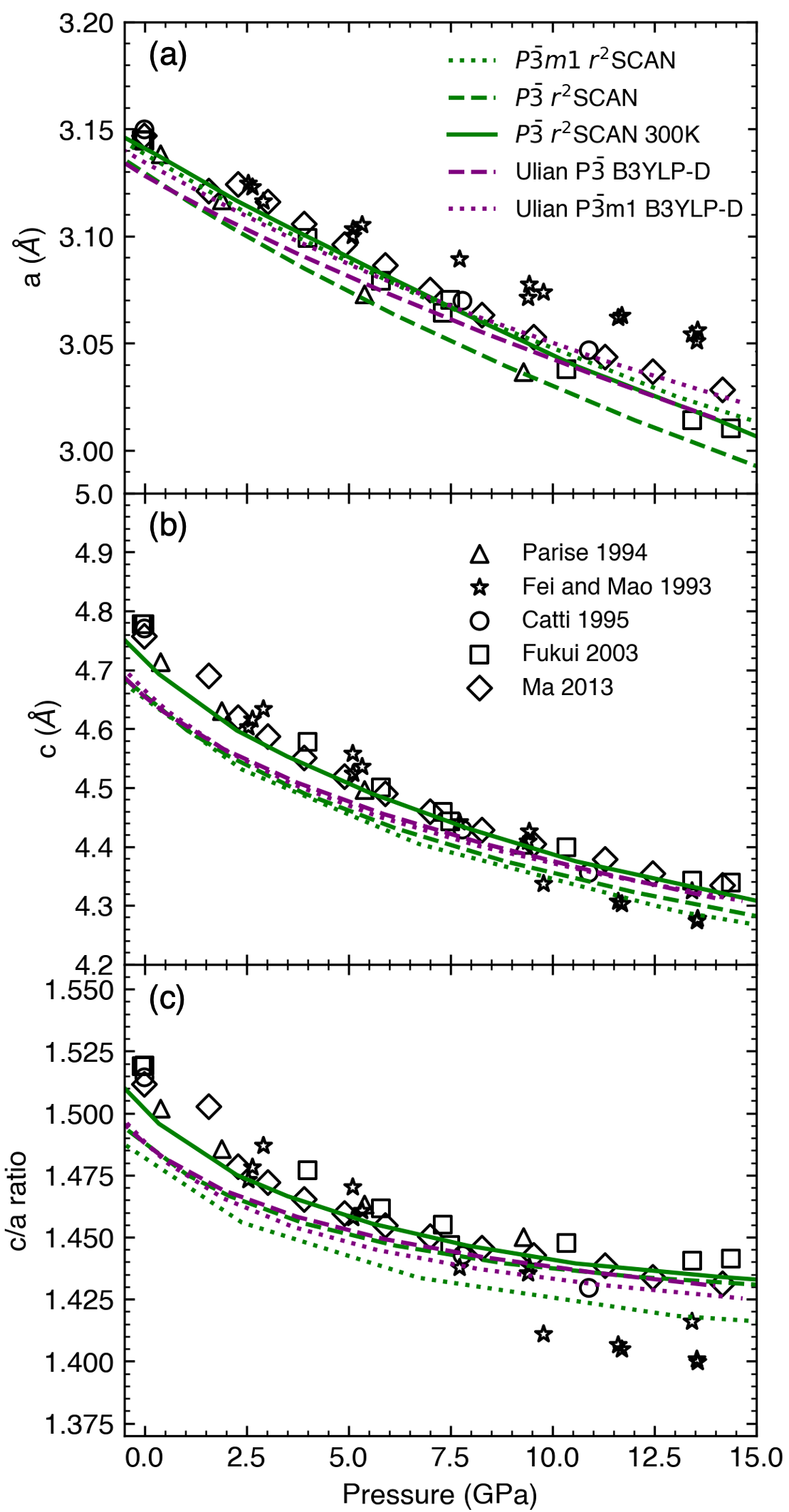}
\caption{Lattice parameters, (a) $a$,  (b) $c$, and (c) $c/a$ ratio in $\pbtm$ and $\pbt$ brucite calculated with $r^2$SCAN. Dashed lines: $\pbt$; dotted lines: $\pbtm$. Open black symbols represent experimental data \cite{parise1994pressure,feimaobrucite,catti1995static,ma2013brucite}. Purple lines are results from previous calculations \cite{ulian2019brucite}.}
\label{lattice_fig}
\end{figure}

\begin{figure}[h]
\centering
\includegraphics[width=0.40\textwidth]{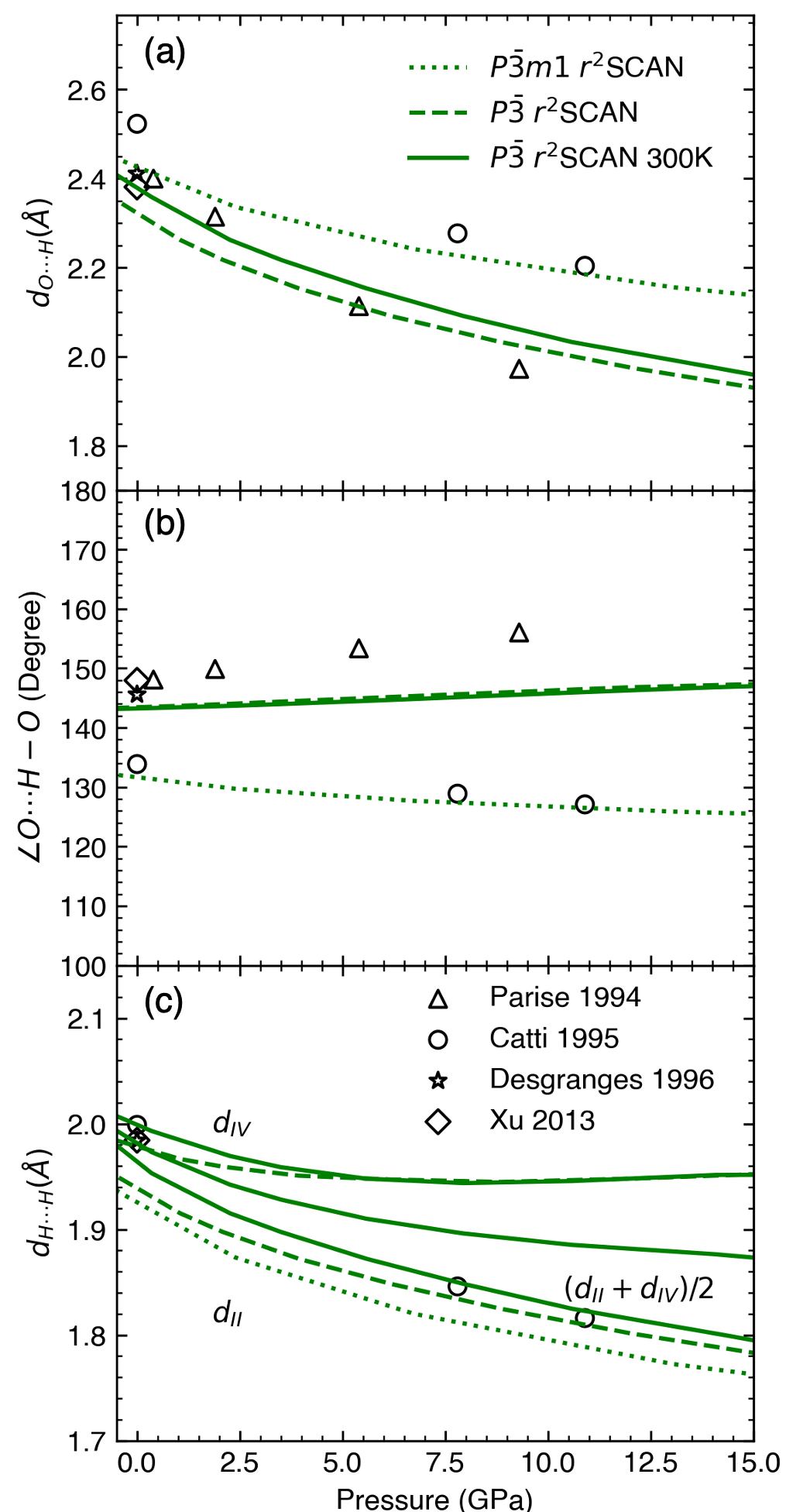}
\caption{Hydrogen bond length (a), angle (b), and $\mathrm{H\cdots H}$ distances (c) in $\pbtm$ and $\pbt$ brucite calculated using $r^2$SCAN and compared with experiments fitted to different symmetries \cite{parise1994pressure,catti1995static,desgranges1996interlayer,xu2013high}. Measurements marked with D represent Mg(OD)$_2$.}
\label{hbond_fig}
\end{figure}
With a reliable EoS, we can investigate the structure evolution under pressure.
Fig.~\ref{lattice_fig} shows the lattice parameters of optimized $P\bar{3}$ and $P\bar{3}m1$ brucite structures obatined with \rscan compared to measurements \cite{parise1994pressure,feimaobrucite,catti1995static,ma2013brucite} and previous B3LYP calculations \cite{ulian2019brucite}.
Experimental results obtained using X-ray diffraction \cite{feimaobrucite,ma2013brucite} and neutron diffraction \cite{parise1994pressure,catti1995static}) are consistent, except for the earliest one \cite{feimaobrucite} which is slightly off.
Compared to experiments, 300~K \rscan calculations give the most consistent description of the lattice parameters and the anisotropic compression behavior described by the $c/a$ ratio.
Both $\pbt$ and $\pbtm$ are more compressible along the $c$-axis, as the structures have weak bonds along this direction.

Interlayer interaction is the result of the two forces: the attractive O$\cdots$H---O forces and the repulsive H$\cdots$H forces.
The accurate description of both types of interactions plays an important role.
Neutron diffraction experiments can reveal the position of light elements (i.e., H) in the structure.
However, whether to fit the structure using $P\bar{3}$ (H at $6i$) or $P\bar{3}m1$ (H at $2d$) symmetry has been a subjective choice.
For instance, proton positions in Ref.~\cite{catti1995static} were fitted to the $2d$-type site, while in Ref.~\cite{parise1994pressure} the $6i$-type site was used.

Figs~\ref{hbond_fig}a and b compare the H-bond length ($d_\mathrm{O{\cdots}H}$) and the bond angle ($\angle\mathrm{O{\cdots}H\hbox{---}O}$) for $P\bar{3}$ and $P\bar{3}m1$ brucite structures obtained in the static and 300~K \rscan calculation.
They also show experimental measurements fitted to these two symmetries.
We notice that the H-bond prediction agrees well with measurements in both phases, especially around 0~GPa. 
Notice that Refs.~\cite{parise1994pressure,xu2013high} reported data on D instead of H, which could change $d_\mathrm{O\cdots H}$ by as much as 0.1~{\AA} in a similar system also rich in H-bonds, i.e., $\delta$-AlOOH \cite{sano2018dalooh}.
The H-bond angle (see Fig.~\ref{fig:structure}b) in $\pbtm$ brucite is in excellent agreement with measurements \cite{catti1995static}.
Although the predicted H-bond angle of $\pbt$ brucite deviates from measurements by 3--5\%, the pressure dependence of the H-bond behavior is correctly captured; for $\pbt$ brucite, H-bonds are shorter, and the $\angle\mathrm{O{\cdots}H\hbox{---}O}$ connecting two layers is closer to 180$^\circ$ compared to those in the $\pbtm$ phase. 
This suggests the H-bond in $\pbt$ structure is stronger than in the $\pbtm$ one.
Under pressure, the H-bond in $\pbt$ shortens and strengthens with increasing H-bond angle.
The opposite happens in the $\pbtm$ structure, i.e., this bond length decreases with decreasing angle, suggesting the H-bond becomes less stable. 

Figure~\ref{hbond_fig}c shows the $\mathrm{H\cdots H}$ distances ($d_\mathrm{H{\cdots}H}$).
As a result of lower symmetry and proton disorder, the $\pbt$ model has two distinct $\mathrm{H\cdots H}$ distances between layers, $d_\mathrm{II}$ and $d_\mathrm{IV}$ \cite{mookherjee2006high}. 
The visualization of $d_\mathrm{II}$ and $d_\mathrm{IV}$ type $\mathrm{H\cdots H}$ is shown in the supplemental Fig.~S3.
They have distinct behavior under pressure: the $d_\mathrm{II}$ decreases monotonically with pressure, while the $d_\mathrm{IV}$ decreases at first and starts to increase at around 7.5~GPa.
This behavior is consistent with previous PBE calculations by Mookherjee and Stixrude \cite{mookherjee2006high}.
The average $d_\mathrm{H\cdots H}$, or $ \tfrac{1}{2} (d_\mathrm{II}$ + $d_\mathrm{IV})$, agrees well with various measurements \cite{parise1994pressure,catti1995static,desgranges1996interlayer,xu2013high} at ambient pressures.
The $d_\mathrm{H\cdots H}$ in $\pbtm$ is much shorter than $\pbt$, indicating a much stronger repulsion.

The stable phonon dispersion we get using \rscan highlights the necessity of including non-local interactions in the brucite system.
As shown in Fig.~\ref{hbond_fig}, the $\pbt$ structure has a relatively stronger attractive H-bond and weaker $\mathrm{H\cdots H}$ repulsive interaction.
The nonlocal interaction captured using \rscan stabilizes the structure and yields stable phonons that would otherwise be unattainable using functionals without nonlocal interactions as captured by this meta-GGA.

\subsection{Elasticity \& velocity}

\begin{figure}[htbp]
\centering
\includegraphics[width=0.45\textwidth]{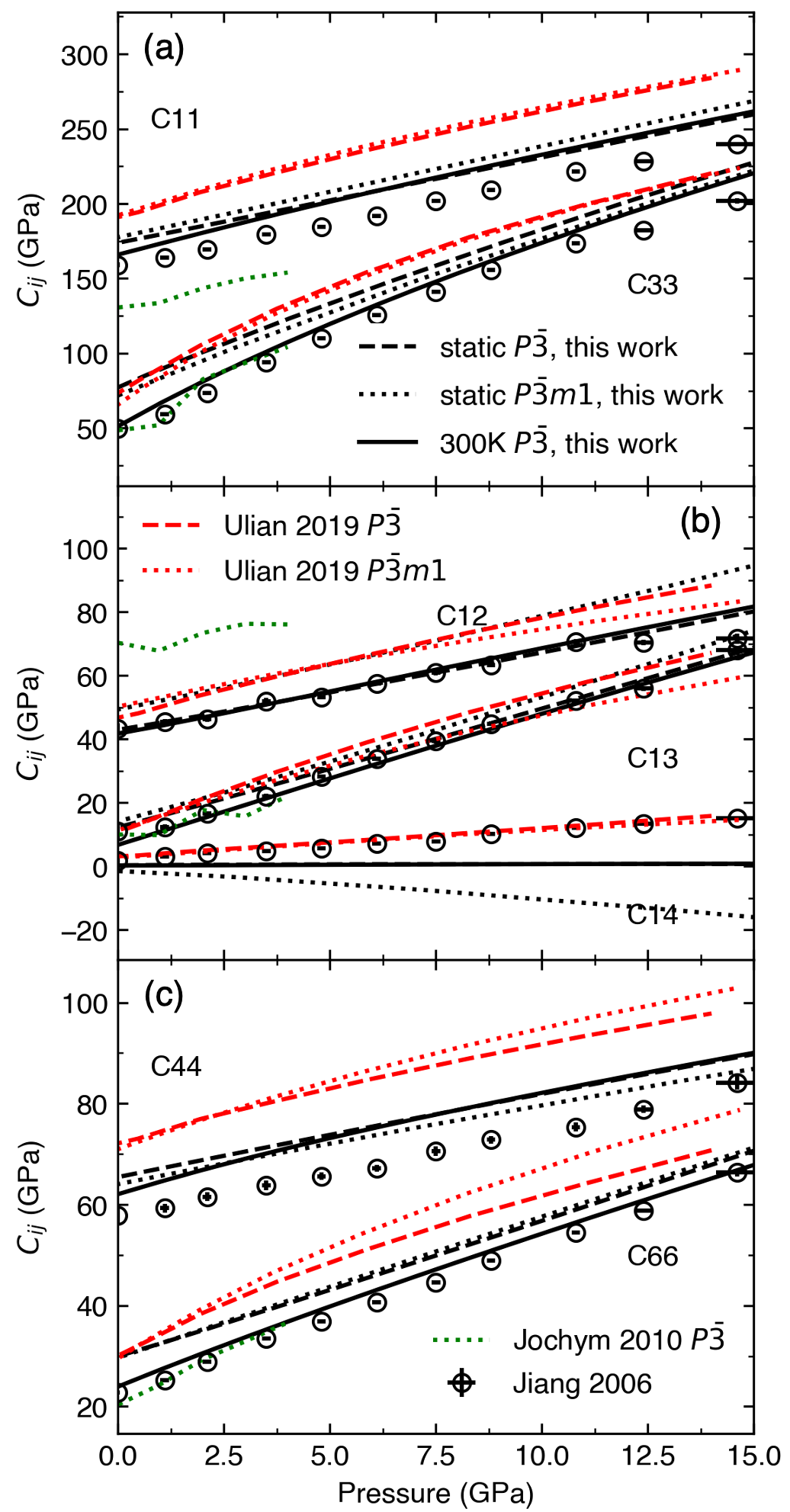}
\caption{Elastic tensor components ($c_{ij}$) vs.\ pressure compared with single crystal measurments \cite{jiang2006brucite} and calculations \cite{jochym2010brucite,ulian2019brucite}.}
\label{elasticity}
\end{figure}

With stable phonons from \rscan calculations, we can compute the 300~K elastic properties and velocities of brucite with the SAM-Cij method \cite{wuQuasiharmonicThermalElasticity2011, luoCijPythonCode2021}.
The numbers of independent elastic tensor terms ($c_{ij}$) of $\pbtm$ and $\pbt$ are slightly different \cite{bruggerPureModesElastic1965}: they both have $c_{11}$, $c_{12}$, $c_{13}$, $c_{14}$, $c_{33}$, $c_{44}$, and $c_{66}=\tfrac{1}{2}(c_{11}-c_{12})$. 
In addtion, $\pbt$ has an independent non-zero $c_{15}$.
Fig.~\ref{elasticity} shows the static and room temperature elastic tensor components, $c_{ij}$, of $\pbtm$ and $\pbt$ compared to measurements \cite{jiang2006brucite} and previous calculations \cite{jochym2010brucite,ulian2019brucite}.
Overall, for both $\pbtm$ and $\pbt$, \rscan results agree better with measurements than PBE \cite{jochym2010brucite} and B3LYP \cite{ulian2019brucite} calculations, except for $c_{14}$.
However, considering the relatively small value of such an elastic constant, we can conclude that \rscan performs well.
Generally, the 300~K results give a correction that decreases the elastic constants and improves agreement with measurements.
For $c_{33}$, the correction can be as large as 30\% at 0~GPa.
Unfortunately, this type of high-temperature elastic constant calculation cannot be carried out for the $\pbtm$ structure because there are unstable phonon; nevertheless, the static results provide insight into the potential differences in elasticity between $\pbtm$ and $\pbt$ brucite.
Although the values are very similar, $\pbt$ brucite agrees better with experiments than $\pbtm$. 
This coincides with our conclusion based on phonon stability.

\begin{figure}[htbp]
\centering
\includegraphics[width=0.45\textwidth]{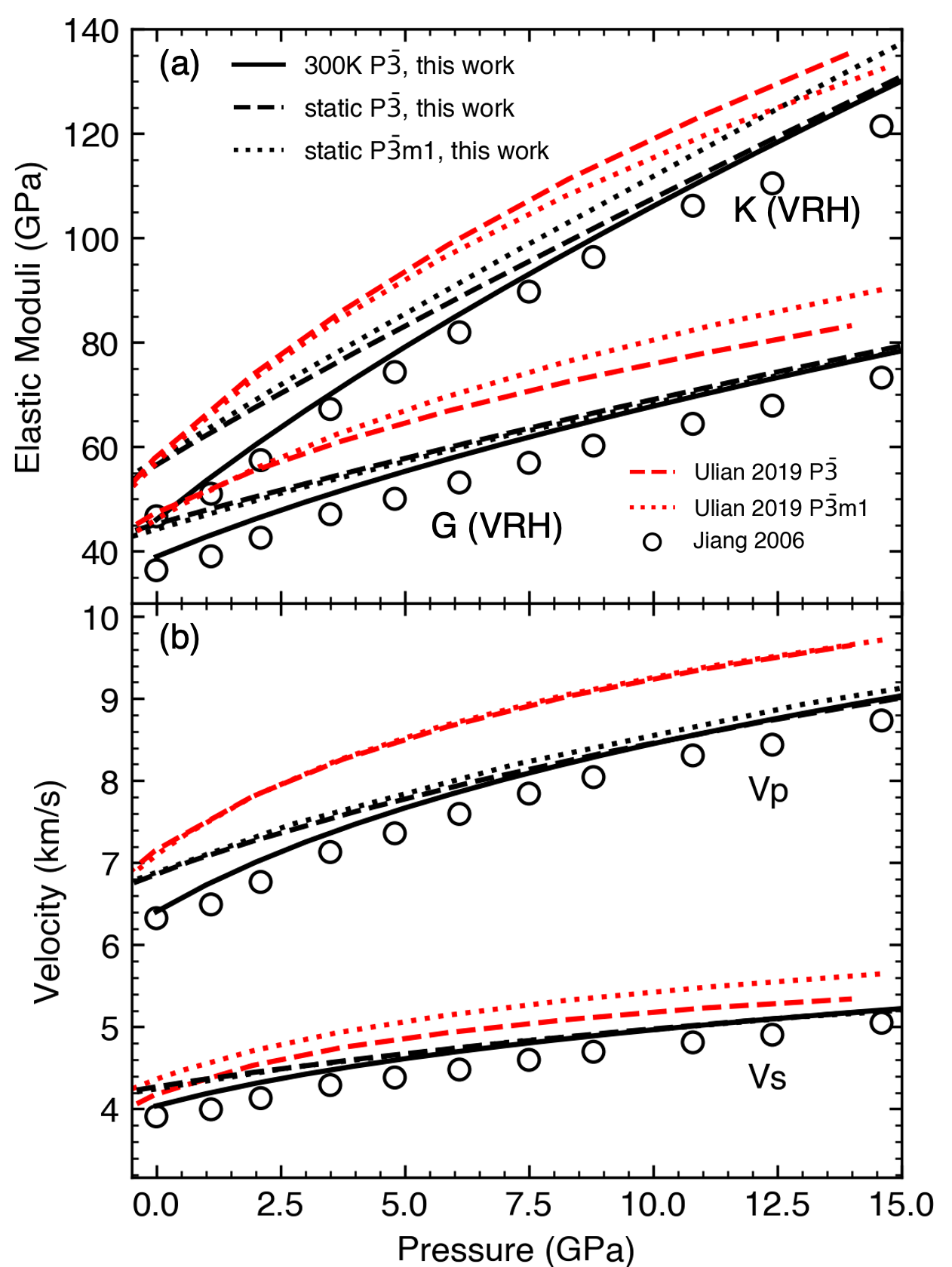}
\caption{
(a)~Elastic moduli and (b)~seismic velocities vs.\ pressure computed at static and 300~K condition compared to previous B3LYP calculation \cite{ulian2019brucite} and measurements \cite{jiang2006brucite}.}
\label{velocity}
\end{figure}

Using elastic constants, we can calculate the Voigt-Reuss-Hill (VRH) average of the bulk and shear moduli.
We compare our results with measurements \cite{jiang2006brucite} and previous calculations \cite{ulian2019brucite} (Fig.~\ref{velocity}a).
Our \rscan results agree better with measurements \cite{jiang2006brucite} than previous B3YLP-D results \cite{ulian2019brucite}.
In particular, our static results for the $\pbt$ structure are closer to measurements than those for the $\pbtm$, and with thermal correction, the 300~K results agree even better at all pressures. 

The static and 300~K longitudinal and transverse velocity ($V_P$ and $V_S$) and their pressure dependency are also plotted and compared with measurements \cite{jiang2006brucite} and previous calculations \cite{ulian2019brucite}.
Note that the measured $V_P$ and $V_S$ are calculated from the data from the original paper \cite{jiang2006brucite} using the following relations: $V_P=\sqrt{(K_\mathrm{VRH}+\tfrac{4}{3}G_\mathrm{VRH})/\rho}$, $V_S=\sqrt{G_\mathrm{VRH}/\rho}$.
In the low pressures, static and 300~K results differ the most. The agreement between our 300~K results and “measurements” is excellent, especially at 0~GPa, where they are almost identical.
This result affirms that \rscan can reproduce the measurements in such a system and that $\pbt$ is the most suitable description of the brucite structure, even at ambient conditions.
The $V_P$ and $V_S$ of $\pbt$ at 300~K, 500~K, and 700~K are listed in Table~SI in the supplementary materials.

\section{Conclusions}
\label{sec:conclusion}

Using \textit{ab initio} calculations, we investigate the stability and elasticity of brucite.
We tested several exchange-correlation functionals, including LDA, PBE, and \rscan. 
The \rscan functional outperforms other functionals (LDA, PBE, PBEsol, and B3LYP) in reproducing the 300~K compression curve and several key structure features involving H-bonds when compared to measurements.
The \rscan functional also gives a dynamically stable $\pbt$ brucite structure at relevant pressures.
Such stable phonon dispersions combined with the QHA are used to compute 300~K elastic properties and acoustic velocities of $\pbt$ brucite; results are in excellent agreement with the measurements.
These successful calculations indicate that $\pbt$ is the suitable space group description of the brucite structure at ambient and elevated pressures.
The success of the \rscan functional for brucite suggests \rscan should also be a suitable choice for studying other sheet hydrous minerals at relevant geophysical conditions of pressure and temperature.

\section*{Acknowledgments}

DOE Award DE-SC0019759 supported this work. Calculations were performed on the Extreme Science and Engineering Discovery Environment (XSEDE) \cite{townsXSEDEAcceleratingScientific2014} supported by the NSF grant \#1548562 and Advanced Cyberinfrastructure Coordination Ecosystem: Services \& Support (ACCESS) program, which is supported by NSF grants \#2138259, \#2138286, \#2138307, \#2137603, and \#2138296 through allocation TG-DMR180081. Specifically, it used the \textit{Bridges-2} system at the Pittsburgh Supercomputing Center (PSC), the \textit{Anvil} system at Purdue University, the \textit{Expanse} system at San Diego Supercomputing Center (SDSC), and the \textit{Delta} system at National Center for Supercomputing Applications (NCSA).

\appendix

\bibliography{brucite}

\end{document}

\normalcolor
\widetext

\setcounter{section}{0}
\renewcommand{\thesection}{S-\Roman{section}}
\setcounter{figure}{0}
\renewcommand{\thefigure}{S\arabic{figure}}
\setcounter{table}{0}
\renewcommand{\thetable}{S\Roman{table}}

\title{\textit{Ab initio} study on the stability and elasticity of brucite
\MakeUppercase{Supplementary Information}
}

\author{Hongjin Wang\,\orcidlink{0009-0008-9766-1177}}
\affiliation{Department of Applied Physics and Applied Mathematics, Columbia University, New York, New York 10027, USA}
\author{Chenxing Luo\,\orcidlink{0000-0003-4116-6851}}
\affiliation{Department of Applied Physics and Applied Mathematics, Columbia University, New York, New York 10027, USA}
\author{Renata M. Wentzcovitch\,\orcidlink{0000-0001-5663-9426}}
\email[]{rmw2150@columbia.edu}
\affiliation{Department of Applied Physics and Applied Mathematics, Columbia University, New York, New York 10027, USA}
\affiliation{Department of Earth and Environmental Sciences, Columbia University, New York, New York 10027, USA}
\affiliation{Lamont--Doherty Earth Observatory, Columbia University, Palisades, New York 10964, USA}
\affiliation{Data Science Institute, Columbia University, New York, NY 10027, USA}

\date{\today}
\maketitle

\textbf{This PDF file includes:}
\begin{itemize}
    \item Figs.~S1 to S3
    \item Table.~S1
\end{itemize}

\begin{figure}[p]
\centering
\includegraphics[scale=0.8]{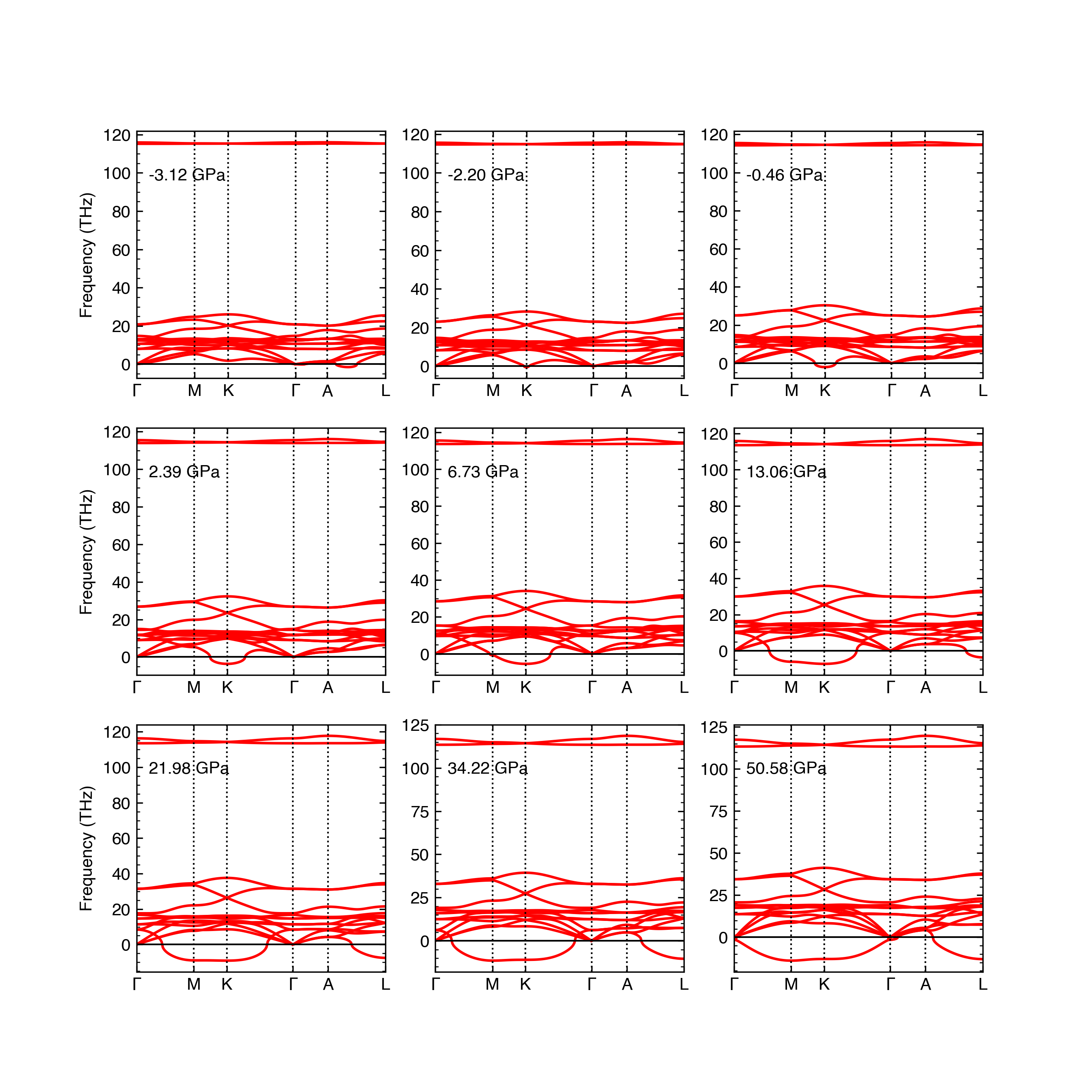}
\caption{Phonon bands of P$\bar{3}m1$ brucite calculated at different pressure ranges}
\label{phonon Pb3m1}
\end{figure}

\begin{figure}[p]
\centering
\includegraphics[scale=0.8]{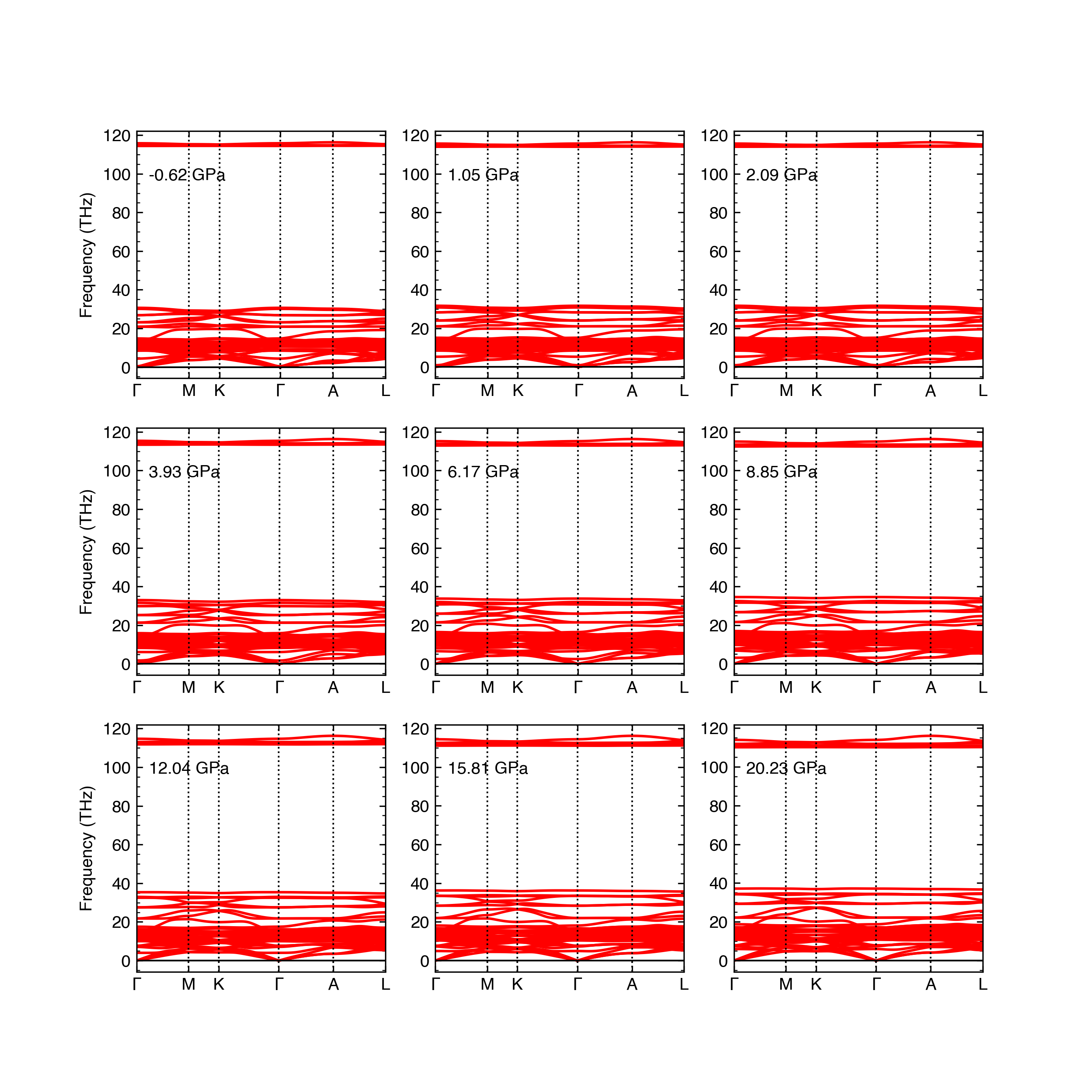}
\caption{Phonon bands of P$\bar{3}$ brucite calculated at different pressure ranges}
\label{phonon Pb3}
\end{figure}
\clearpage

\begin{figure}[p]
\centering
\includegraphics[scale=0.8]{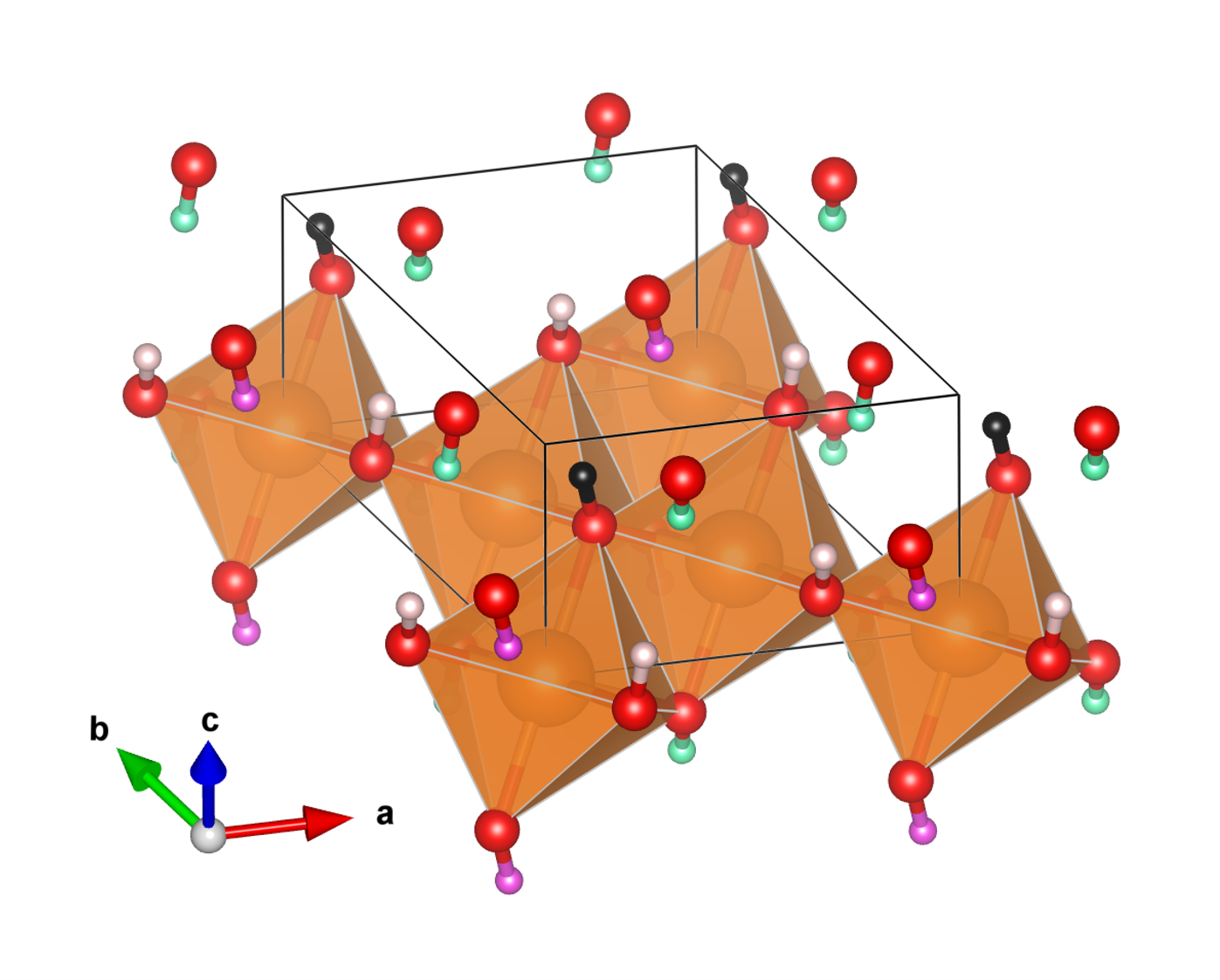}
\caption{Visualization of $d_\mathrm{II}$ (black and green) and $d_\mathrm{IV}$ (black and pink) type $\mathrm{H\cdots H}$}
\label{phonon Pb3}
\end{figure}
\clearpage

% \bibliography{Geophysics}
\begin{table*}
\renewcommand{\arraystretch}{1.25}
\small
\caption{Longitude and transverse sound velocity of $P\bar{3}$ brucite at 300, 500, and 700 Kelvin.}
\label{structure_table}
\begin{ruledtabular}
\begin{tabular}{ccccccc}
%\toprule
P (GPa) &
\begin{minipage}{5em}\centering $V_P$, 300~K \\(km/s)\end{minipage} &
\begin{minipage}{5em}\centering $V_S$, 300~K \\(km/s)\end{minipage} &
\begin{minipage}{5em}\centering $V_P$, 500~K \\(km/s)\end{minipage} &
\begin{minipage}{5em}\centering $V_S$, 500~K \\(km/s)\end{minipage} &
\begin{minipage}{5em}\centering $V_P$, 700~K \\(km/s)\end{minipage} &
\begin{minipage}{5em}\centering $V_S$, 700~K \\(km/s)\end{minipage} \\
\hline
0.0 & 6.4032 & 6.1123 & 5.7696 & 4.0372 & 3.8733 & 3.7352 \\
1.0 & 6.7293 & 6.5208 & 6.2352 & 4.1856 & 4.0660 & 3.9048 \\
2.0 & 7.0074 & 6.8545 & 6.6428 & 4.3122 & 4.2224 & 4.1004 \\
3.0 & 7.2524 & 7.1352 & 6.9755 & 4.4237 & 4.3530 & 4.2592 \\
4.0 & 7.4699 & 7.3805 & 7.2586 & 4.5225 & 4.4667 & 4.3934 \\
5.0 & 7.6674 & 7.5992 & 7.5039 & 4.6120 & 4.5677 & 4.5089 \\
6.0 & 7.8477 & 7.7966 & 7.7239 & 4.6935 & 4.6585 & 4.6121 \\
7.0 & 8.0145 & 7.9775 & 7.9215 & 4.7686 & 4.7413 & 4.7042 \\
8.0 & 8.1692 & 8.1443 & 8.1035 & 4.8381 & 4.8174 & 4.7886 \\
9.0 & 8.3144 & 8.2995 & 8.2709 & 4.9031 & 4.8879 & 4.8659 \\
10.0 & 8.4507 & 8.4450 & 8.4270 & 4.9639 & 4.9538 & 4.9376 \\
11.0 & 8.5794 & 8.5817 & 8.5734 & 5.0213 & 5.0154 & 5.0045 \\
12.0 & 8.7017 & 8.7112 & 8.7109 & 5.0755 & 5.0736 & 5.0671 \\
13.0 & 8.8178 & 8.8342 & 8.8416 & 5.1270 & 5.1286 & 5.1264 \\
14.0 & 8.9287 & 8.9510 & 8.9655 & 5.1760 & 5.1808 & 5.1824 \\
15.0 & 9.0348 & 9.0630 & 9.0835 & 5.2228 & 5.2306 & 5.2355 \\
%\bottomrule
\end{tabular}
\end{ruledtabular}
\end{table*}
\clearpage